\documentclass[twoside,onecolumn]{article}
\usepackage{hyperref}
\usepackage{apacdoc} 
\usepackage[sc]{mathpazo} 
\usepackage[T1]{fontenc} 
\usepackage{microtype} 

\usepackage[english]{babel} 

\usepackage[hmarginratio=1:1,top=32mm,columnsep=20pt]{geometry} 
\usepackage[hang, small,labelfont=bf,up,textfont=it,up]{caption} 
\usepackage{booktabs} 

\usepackage{lettrine} 

\usepackage{enumitem} 
\setlist[itemize]{noitemsep} 

\usepackage{abstract} 
\usepackage{titlesec} 
\renewcommand \thesection{\arabic{section}}
\renewcommand \thesubsection{\arabic{section}.\arabic{subsection}}
\renewcommand \thesubsubsection{\arabic{section}.\arabic{subsection}.\arabic{subsubsection}}

\titleformat{\section}[block]{\large\scshape\centering}{\thesection.}{1em}{} 
\titleformat{\subsection}[block]{\normalfont\bfseries\normalsize}{\thesubsection.}{1em}{} 
\titleformat{\subsubsection}[block]{\normalfont\itshape\normalsize}{\thesubsubsection.}{1em}{} 

\usepackage{fancyhdr} 
\usepackage{titling} 

\usepackage{hyperref} 

\pretitle{\begin{center}\huge\bfseries} 
\posttitle{\end{center}} 
\title{Reimagining Near-Earth Space Policy in a Post-COVID World} 
\author{%
\\
\textsc{John C.~Barentine}\\[1ex] 
\small Dark Sky Consulting, LLC\\
\small 9420 E Golf Links Rd Ste 108 PMB 237, Tucson, AZ 85730-1317 USA\\ 
\small \href{mailto:john@darkskyconsulting.com}{john@darkskyconsulting.com}\\
\\
\textsc{Jessica Heim}\\[1ex] 
\small University of Southern Queensland Centre for Astrophysics \\
\small West Street, Toowoomba Qld 4350, Australia\\
\small \href{mailto:Jessica.Heim@usq.edu.au}{Jessica.Heim@usq.edu.au}\\
\\
\textsc{Aparna Venkatesan}\\[1ex] 
\small University of San Francisco Department of Physics and Astronomy\\
\small 2130 Fulton Street, San Francisco, CA 94117 USA\\ 
\small \href{mailto:avenkatesan@usfca.edu}{avenkatesan@usfca.edu}\\
\\
\textsc{James Lowenthal}\\[1ex] 
\small Five College Astronomy Department \\
\small Smith College, 44 College Lane, Northampton, MA 01063 USA\\ 
\small \href{mailto:jlowenth@smith.edu}{jlowenth@smith.edu}\\
\\
\textsc{Monica Vidaurri}\\[1ex] 
\small Stanford University School of Earth, Energy \& Environmental Sciences,\\
\small 397 Panama Mall, Stanford, CA 94305-2210 USA\\ 
\small \href{mailto:mvidaurri@stanford.edu}{mvidaurri@stanford.edu}
}
\date{} 
\renewcommand{\maketitlehookd}{%
\begin{abstract}
\noindent Our planet and our species are at an existential crossroads. In the long term, climate change threatens to upend life as we know it, while the ongoing COVID-19 pandemic revealed that the world is unprepared and ill-equipped to handle acute shocks to its many systems. These shocks exacerbate the inequities and challenges already present prior to COVID in ways that are still evolving in unpredictable directions. As weary nations look toward a post-COVID world, we draw attention to both the injustice and many impacts of the quiet occupation of near-Earth space, which has rapidly escalated during this time of global crisis. The communities most impacted by climate change, the ongoing pandemic, and systemic racism are those whose voices are missing as stakeholders both on the ground and in space. We argue that significant domestic and international changes to the use of near-Earth space are urgently needed to preserve access to — and the future utility of — the valuable natural resources of space and our shared skies. After examining the failure of the U.S. and international space policy status quo to address these issues, we make specific recommendations in support of safer and more equitable uses of near-Earth space.
\end{abstract}
}

\begin{document}\sloppy

\maketitle


\section{Introduction and background}

\subsection{Uses of low-Earth orbit space 1957-2019}
Direct human involvement in outer space began with the launch of the Soviet Sputnik 1 satellite on October 4, 1957, although objects were launched above the Kármán line -- one definition of the limit of outer space -- as early as 1944. The United States responded quickly, launching its first satellite, Explorer 1, on 1 February 1958. Near-Earth orbital space was quickly exploited for telecommunications purposes beginning in 1958 with the U.S. Project SCORE (Signal Communications by Orbiting Relay Equipment), the first purpose-built communications satellite. Passive, long-distance communications were achieved in 1960 using large balloons inflated after reaching orbit. 

Designs for flying satellites in groups with similar orbital characteristics, called ``constellations,'' were first realized for civilian use with the deployment of the initial Global Positioning System (GPS) satellite in 1978. Long-term human habitation in low-Earth orbit (LEO) began in 1971 with the launch of the Soviet Salyut 1 space station. Small and inexpensive ``CubeSats'' were first launched at the turn of the 21st century. As of late 2021, there were about 29,000 objects larger than 10 centimeters in size orbiting the Earth (ESA, 2021), of which at least 4,550 were operational satellites (UCS, 2021).

\subsection{Arrival of the satellite 'megaconstellation' era}
Until recently, satellite constellations tended to comprise no more than a few dozen objects. An example of this is the GPS constellation, consisting of 24 satellites in six orbital planes at an altitude of 20,180 km GPS satellites are therefore slow in their angular rate of motion across the sky, allowing for multiple objects to remain above the horizon at any given location on the Earth. Navigational and communications constellations are usually placed at medium-Earth orbit (MEO) altitudes to achieve a similar effect.

A new era in the use of LEO space began in May 2019 with the first launch of objects in SpaceX's 'Starlink' satellite system, a planned constellation of 42,000 communications satellites intended to provide low-latency broadband service with near-global coverage (Freeman, 2020). Among the stated purposes of SpaceX and its competitors in entering this market is to extend affordable broadband access to unserved and underserved populations, an increasingly pressing issue given the ongoing COVID-19 pandemic and the widening digital divide (Milazzo et al., 2021). However, this premise has been questioned (Rawls et al., 2020). Some authors highlight concerns about the future of Internet governance given the outsized influence of the home countries of launching companies and their potential to control global information flows (Voelsen, 2021).

The start of the Starlink project also brought an unprecedented change in the appearance of the night sky. Bright ``trains'' consisting of dozens of bright satellites moving together as a group were seen from the ground in the days and weeks after each launch (Hall, 2019). These objects soon began to negatively impact the observations of both professional and amateur astronomers as well as astrophotographers (Witze, 2020). Starlink also spawned multiple online petitions seeking interventions by both U.S. regulators and the international community.~\footnote{Petitions open as of December 2021 are available on \href{https://www.change.org/p/stop-spacex-starlink-from-spoiling-outer-space-for-humanity}{change.org}, \href{https://sign.moveon.org/petitions/stop-elon-musk-s-starlink-satellite-program}{MoveOn.org} and \href{https://secure.avaaz.org/community_petitions/en/SpaceX_and_companies_planning_to_launch_constellations_of_satellites_Losing_the_night_sky_due_to_tens_of_thousands_of_satellites/}{Avaaz}.}

The astronomical community is highly concerned about the increasing material harm to ground- and space-based astronomy enterprises. They have begun to engage in constructive dialogue with satellite companies, which led to certain design modifications by the industry intended to render the satellites less visible from the ground. These mitigation efforts have yielded mixed results (Cole, 2020; Tregloan-Reed et al., 2020). Meanwhile, more launches took place in 2021 than in any other year since the beginning of the Space Age (Impey, 2021). Based on published reports, Federal Communications Commission (FCC) filings and other sources, we count over 422,000 LEO objects proposed for deployment since 2019, of which over 2,000 have been launched.

\subsection{Known and anticipated impacts of LEO satellite mega-constellations}
As LEO space begins to fill rapidly with functional and inoperable satellites, discarded space hardware, and pieces of debris (`space objects'), risks to various ground- and space-based activities quickly rise. Boley and Byers (2021) reviewed these risks, identifying ``multiple … tragedies to ground-based astronomy, Earth orbit, and Earth's upper atmosphere.'' We briefly review these concerns below.

\subsubsection{Debris generation}
The probability of close approaches, or 'conjunctions', between space objects in the vicinity of Earth is a function of many variables, including the volume density of objects and the parameters of their orbits. When the distance between objects in conjunctions is less than the physical size of the objects, collisions result. The severity of collisions and the subsequent number of pieces of orbital debris generated depend on factors such as the masses of the objects and their relative speeds at impact.

Collisions can be accidental, such as the February 10, 2009 collision between the active commercial Iridium 33 satellite and the defunct Russian military satellite Kosmos 2251 that generated nearly 2,300 pieces of debris of sufficient size to be tracked from the ground. While collisions between active objects are rare, debris-satellite collisions are more likely and debris-debris collisions are very common (Le May et al., 2018). 

Intentional destruction of satellites is rarer still. To cite a recent example, on November 15, 2021 the Russian Defense Ministry deliberately destroyed the defunct Soviet Kosmos 1408 satellite in a test of its antisatellite (ASAT) weapon capability (Balmforth, 2021). The test generated a field of at least 1,500 trackable debris objects in LEO orbits that threaten the safety of both space operations and human spaceflight (ACA, 2021). Because Kosmos 1408 was in a polar orbit, and its fragments spread out over altitudes ranging from 300 to 1000 km, that debris could potentially collide with any object in LEO, including the International Space Station and the Chinese Tiangong space station. Because of this test, astronauts on board the International Space Station had to take precautionary measures in case of debris impact by donning their space suits and taking shelter (Roulette, 2021). In addition to Russia, the United States, India, and China have all conducted debris-generating ASAT tests since 1985. For a future scenario involving 65,000 satellites in LEO, Thiele and Boley (2021) found that the probability of one or more satellites colliding with ASAT fragments less than one cm in size is roughly 30\% for a single ASAT test.

The potential for collisions to threaten spacecraft in LEO increases as the number of objects in LEO increases. Such instances may not only generate debris through collisions, but they can also threaten the safety of astronauts aboard crewed spacecraft. In late 2021, China reported two incidents in which its Tiangong space station took evasive action to prevent colliding with objects in the SpaceX Starlink constellation (Kwan, 2021).~\footnote{The incidents were reported to the United Nations Office for Outer Space Affairs in “Note verbale dated 3 December 2021 from the Permanent Mission of China to the United Nations (Vienna) addressed to the Secretary-General,” (A/AC.105/1262, 3 December 2021).} Eventually, the consequences of these collisions may become lethal (Pelton et al., 2020).

Orbital dynamics experts have long noted the potential for such debris-generating events to yield cascades of further collisions (Kessler and Cour-Palais, 1978), potentially leaving LEO space inaccessible. We emphasize that the challenge of modeling such nonlinear, unpredictable events positions us in the primarily reactive stance of managing the consequences of collisions after the fact. It is therefore far better to avoid the conditions that could precipitate collisions and to develop systematic reporting and tracking of space objects (Jah, 2020).~\footnote{See also \href{https://www.commerce.senate.gov/services/files/4CE62711-708C-493C-A325-E0B845B34372}{Jim Bridenstine Testimony, Senate Space and Science Subcommittee Hearing, October 21, 2021}.}

\subsubsection{Optical and infrared astronomy}
Because space objects can both directly reflect sunlight and passively emit infrared radiation, they become moving sources of optical and infrared (OIR) light in the night sky. This light may be detected by cameras and telescopes, competing with cosmic light and rendering astronomical images diminished or even useless. 

The extent to which satellites interfere with OIR imaging depends principally on their orbital altitudes, the hour of night, the time of year, the location of the observer on Earth, and the size of the field of view of a camera or telescope (Bassa et al., 2022). The most significant impacts are expected in the twilight hours around sunset and sunrise when the Sun remains above the horizon for objects in LEO orbits while the night sky seen from the ground below grows gradually darker. Hainaut and Williams (2020) found that for the Vera Rubin Observatory, between 30\% and 40\% of its sky survey images may be compromised during these times. Trails of light attributable to Starlink satellites have also appeared in Hubble Space Telescope (HST) images (Ingraham, 2021). A search of the HST data archive using machine learning found more than 2,400 satellite trails in images, and it is expected that the mega-constellation era will see a large increase in trails (Martin et al., 2021).

As the latitude of the observer increases, so does the number of satellites that remain in sunlight at LEO altitudes. If the majority of planned LEO mega-constellations are built, hundreds of satellites may be visible to the unaided eye in twilight as seen during the summer months from latitudes at and above 50° (McDowell, 2020; Lawler et al., 2021). Conversely, at low latitudes and in the middle of the night, few or no satellites will be  visible.

The combined effect of reflected sunlight from myriad space objects can elevate the brightness of the night sky itself. Kocifaj et al.~(2021) found that even before the launch of the first Starlink satellites, space objects already contributed enough diffuse light to raise the brightness of the night sky approximately 10\% above an assumed natural background level. If the population of debris objects increases as expected in the coming decade, then the number of LEO objects in 2030 will be a factor of about 25 times higher than it is now. That could potentially raise the brightness of the night sky seen over much of the world to a level comparable to that seen in and near cities with moderate amounts of light pollution. Unlike individual satellite streaks, this effect cannot be removed with software; it is an omnipresent phenomenon no matter how remote or high the viewing location, impacting astronomical deep-sky exposures, astrophotographers, night-sky tourism, cultural sky traditions and more.

\subsubsection{Radio-frequency interference}
Communications satellites serve as relays of information between widely separated points on the ground. They serve this purpose by receiving radio transmissions from the ground and rebroadcasting them in other directions. Their function is therefore directly enabled by the ability to not only receive but also to emit radio energy, which has the potential to cause radio-frequency interference (RFI).

RFI is more regulated than other aspects of satellite operations and has been for decades. Almost a century ago, the world identified the radio spectrum as a shared resource requiring international coordination to properly manage for the benefit of all. The International Telecommunication Union Radiocommunication Sector (ITU-R) is the United Nations-recognized body that manages both the international radio-frequency spectrum and allocation of satellite orbits. Its mission is ``to ensure the rational, equitable, efficient and economical use of the radio-frequency spectrum by all radiocommunication services, including those using satellite orbits'' (ITU-R, 2021). The modern telecommunications network depends crucially on controlling and limiting RFI in order to ensure the trouble-free operation of its radio links. 

Radio astronomers have long sought to locate their telescopes in isolated, rural ``radio-quiet zones''' to limit the influence of RFI. Satellites disrupt this approach because essentially nowhere in the world is shielded from their influence. It remains a problem even when satellite transmissions are limited to narrow beams, as radio telescopes can receive satellite transmissions from multiple directions. Given their exquisite sensitivity to radio energy, the instruments attached to such telescopes are highly vulnerable to radio energy emitted by satellites, which can damage or even destroy their delicate detectors. 

Even in better circumstances, satellite radio transmissions can have severe impacts on radio astronomy, rendering some observations extremely difficult, if not impossible (Combrinck et al., 1994). Certain parts of the radio spectrum are therefore reserved for radio astronomy access in order to ensure that radio astronomers can continue to use them for purposes of scientific research. Transmissions from the increasing number of communications and navigation satellites orbiting the Earth run the risk of closing these radio ``windows'' on the cosmos even under conditions of proper radio spectrum management.

\subsubsection{Additional stakeholders' engagement with the night sky}
The night sky is an extension of the Earth. It occupies half of one's field of view at night and is often understood as equally a part of the environment as the land beneath one's feet. For cultures around the world, from prehistory to the present, observation of the celestial bodies and their movements are integrated into many facets of daily life including timekeeping and navigation as well as spiritual and religious practices. In many Indigenous cultures, traditional knowledge and ceremonies often have connections to the stars and the concept of relationships between people, the Earth and the stars is often at the heart of these teachings (Lee et al., 2020; Venkatesan et al., 2019). As noted earlier, these real-time observations cannot be software-corrected for satellites and are often conducted toward the horizon or circumpolar skies where such satellites cause most interference. Hamacher et al.~(2020) argue that this "whitewashing of the night sky through colonial policy" is a form of "ongoing cultural and ecological genocide" that can "erase Indigenous connection to the stars." 

Some people simply find inspiration in the night sky, whether through quiet contemplation or the creation of art, storytelling, literature and music. Amateur astronomers access the night sky for recreation and sometimes contribute to astronomical research, while astrophotographers use it as a backdrop for aesthetic representations of the nighttime environment. When the night sky is altered by human activities, connections between people and the cosmos are disrupted. The addition of dozens or hundreds of artificial objects moving across the night sky exacerbates such effects, serving as an unwelcome reminder of the ever-expanding reach of technology and capitalism. Yet, such concerns have largely been ignored in the ongoing debate about how orbital space should be used.

The Community Engagement Working Group of the recent Satellite Constellations 2 (SATCON2) workshop sought to hear perspectives from diverse constituencies, including individuals from a number of Indigenous tribes and communities (Venkatesan et al., 2021). Though many people in such communities may desire increased broadband access, it is clear that there are also concerns about issues of sovereignty and questions as to whether satellite-based internet access is necessarily the best option for a given community's needs. In addition, mega-constellations may interrupt ceremonial engagement with the stars; erasing the stars is akin to erasing Indigenous identities and storytelling in what may be a new form of colonization. The resulting recommendations of the Working Group included the need for dialogue, building long-term relationships with all impacted constituencies, and the "duty to consult" with sovereign Indigenous nations regarding space activities. 

\subsubsection{Air and noise pollution from rocket launches}
Satellite mega-constellations are launched into orbit by conventional rockets. Depending on the type of fuel the rocket burns, pollutants produced during launch can include the greenhouse gas carbon dioxide; large quantities of water vapor released into the upper atmosphere, where it exacerbates depletion of the ozone layer; black carbon particulates; and other harmful materials (Boley and Byers, 2021; Dallas et al., 2020). 

Rockets are launched from facilities that may be located in or near environmentally sensitive areas and/or human communities. The negative effects of frequent high-decibel noise (Tran et al., 2018) and air pollution (Jindal et al., 2018) on those human and natural neighbors can be severe. As the development of LEO mega-constellations accelerates, new ``spaceports'' from which to launch them are being proposed and built around the world, often in the face of strong local opposition (Solon, 2021).

\subsubsection{Ecological impacts of satellite mega-constellations}
Virtually all life on Earth evolved in the regular cycle of bright days and dark nights. Most organisms are subject to circadian rhythms of behavior, activity, metabolism, and hormone production that are controlled at least in part by those natural light and dark patterns (Foster, 2021; LeGates et al., 2014). Until the first Starlink satellites were launched, satellites bright enough to be visible to the unaided human eye were seen so infrequently that they were generally considered inconsequential for living things. The advent of LEO satellite mega-constellations may represent a new potential threat to their well-being. The circadian rhythms of many organisms are apparently controlled by low levels of diffuse light integrated over wide angles of sky and relatively long time periods (Brown, 2016). While individual bright satellites moving across the sky may not interfere with those natural functions, an artificially brightened sky could interfere with them very strongly.

Large numbers of individual satellites could also interfere with seasonal migration and navigation. A wide variety of species from birds to mammals to insects have been shown to navigate at least partly using the stars and the Milky Way (Dacke et al., 2021; Emlen, 1970; Mauck et al., 2008; Mouritsen and Larsen, 2001). Effects on these natural behaviors when the real stars are surrounded or overwhelmed by large numbers of bright, moving ``fake stars'' remain to be seen. The problem is so new that results of scientific studies have yet to appear in the peer-reviewed literature, but concerns are beginning to surface (Lintott and Lintott, 2020). The precautionary principle dictates that reasonable efforts should be made to understand and mitigate the problem before irreversible damage is done.

\subsubsection{Pollution of the upper atmosphere and oceans from satellite re-entry}
Lastly, reentering space hardware currently represents a return of mass from orbit approaching 100 metric tons per year (Pardini and Anselmo, 2019). Ablation and destruction of these materials during decommissioning and reentry is expected to deposit significant metals in the upper atmosphere, greatly exceeding the contribution from the natural flux of micrometeoroids (Schulz and Glassmeier, 2021). It is not currently known what impact this will have on the chemistry or radiative equilibrium of the upper atmosphere; as a point of reference, the deliberate injection of aluminum at such altitudes has been proposed as a means of changing the Earth's reflectivity to sunlight (Keith, 2000). Given current uncertainties, the proliferation of objects in LEO and their potential to reenter the atmosphere amounts to a complex, uncontrolled geoengineering experiment—one that has a significant chance of becoming irrevocably intertwined with climate change.

There are also myriad, and equally unknown, collateral effects. For instance, particulate matter resulting from satellite re-entry will eventually fall to the Earth's surface, mostly in ocean water, adding more pollutants to already stressed ecosystems (De Lucia and Iavicoli, 2019). And re-entering satellites and debris that contribute substantial sodium to the upper atmosphere may raise the brightness of the airglow layer at an altitude of 100 km(Rosenberg, 1966). In a steady state of re-entries, large satellite constellations may contribute enough sodium to permanently enhance the airglow, diminishing the visibility of stars from the ground. The situation awaits systematic studies and modeling to determine how significant this problem may be for astronomy.


\section{Acute inadequacy of the policy status quo}

Without changes to existing space policies, the current and potential impact of satellite mega-constellations on human endeavors and the environment is immense in scope and profound in effect. Importantly, the issues discussed are only those that are currently anticipated and have begun to attract research interest. However, unexpected developments will likely arise.

The advent of satellite mega-constellations does not merely represent a predictable scaling up of concerns that existed in the preceding space era. Rather, due to the sheer magnitude of objects being launched, the rapidly growing space sector is creating entirely new problems, while simultaneously exacerbating existing issues. Orbital crowding raises collision risks exponentially (Rossi et al., 1994), while for the first time, astronomers face the prospect of significant data loss due to satellite trails fouling their images (Martin et al., 2021). 

History provides many instances of the pace of technology exceeding human understanding of its full implications and leading to negative externalities (Nye, 2006). Unequal power relationships and competing views often result in those groups with more social and political power prevailing in disputes about how technology is used. The history of the use of outer space is dominated by the powerful, while the marginalized are generally ignored (Dickens and Ormrod, 2007). The current, billionaire-funded space race only adds to this historical trend, with these companies showing little interest in, and even active disdain for, regulatory efforts and the safety and working conditions of their employees in the Earthly domain (e.g., Siddiqui, 2021), boding ill for their space initiatives.

The U.S. National Science Foundation and United Nations Office for Outer Space Affairs have recently convened four conferences of experts to better understand the potential harms associated with mega-constellations and to seek potential solutions (Walker et al., 2020a; Green et al., 2020; Hall et al., 2021). In parallel, they examined the current world policy regime and found that existing policies are limited in their ability to confront and address arising concerns. While the analysis of the status quo remains in process, we argue that not only are current policies fundamentally inadequate to deal with the new reality that megaconstellations represent, but also that their inadequacy is fueling a ``Wild West'' race to occupy LEO.

\subsection{The current international and U.S. policy regimes}
International space policy largely descends from the Outer Space Treaty (OST, 1967). The OST’s main provisions include blanket prohibitions on the deployment of nuclear weapons and making territorial claims in space, a demand that space is for the exploration and use of all nations equally, and limits on those uses to peaceful purposes. It permits some military activities in space but enjoins against unrestricted competition for resources that could result in armed conflict. Parties to the OST are required to implement its provisions through national laws. Article VI of the OST requires that state parties to the Treaty ``bear international responsibility for national activities in outer space, … whether such activities are carried on by governmental agencies or by non-governmental entities.'' Responsibility is enforced through the accompanying Space Liability Convention (1972). 

However, there are many topics on which the OST is entirely silent, such as the extraction of natural resources from Solar System bodies. There is little in the way of formal international regulation and instead more reliance on adherence to uncodified best practice (Rotola and Williams, 2021). Consensus-based policymaking at such high levels generally proceeds at a glacial pace and cannot keep up with the rapid evolution of technology that has led to uses of space undreamt of by the framers of the OST.

Because most major commercial satellite operators currently launch from U.S. territory, its policies de facto govern the activities of many entities in space. The Federal Aviation Administration (FAA) regulates launch activities, such as securing airspace, but its legal jurisdiction ends before objects reach ``space,'' which it defines informally to be an altitude of 80 kilometers (Gouyon Matignon, 2020).~\footnote{14 CFR Chapter III.} For activities in space involving radio communication with the ground, the Federal Communications Commission (FCC) licenses operations. It further requires that satellite operators specify plans to avoid creation of space debris through responsible post-mission disposal.~\footnote{85 FR 52422.}

The National Environmental Policy Act (NEPA) of 1969 does not expressly prohibit the FCC from licensing radio transmissions from space to ground, but it does insist that the FCC fairly examine all of the environmental impacts of doing so. However, the FCC refuses to fully implement NEPA, and it unilaterally imposes a ‘categorical exemption’ on virtually all satellite operations from environmental impact review (Ryan, 2020). Furthermore, the FCC does not consider night-sky impacts in its decisions to license satellite operations. 

FCC resistance to the authority of NEPA has come under scrutiny in the federal courts.~\footnote{For instance, in 2018, the FCC changed its rules to eliminate the application of NEPA, as well as the National Historic Preservation Act, to its authorization of small-cell networks increasingly relied upon to provide 5G services across the country. 16 Indian Nations sued the FCC, and a federal court found the FCC’s action unlawful (\emph{United Keetoowah Band of Cherokee Indians in Oklahoma v. FCC}, No. 18-1129, D.C. Cir. 2019).} Viasat recently sued the FCC in the D.C. Circuit to compel it to apply NEPA review to SpaceX launches, but the case remains pending.~\footnote{\emph{Viasat, Inc., v. FCC and Space Exploration Holdings, LLC}, No. 21-1123 (D.C. Cir. 2021).} Gilbert and Vidaurri (2021) examined existing national and international case law, concluding that NEPA should be applied generally to space activities operating under U.S. jurisdiction.	 

\subsection{Policy implications of LEO space as a global commons}
At the heart of this issue is the matter of how the use of outer space is, and should be, governed. Space is not obviously the province or territory of any one nation. As such, we can look to terrestrial analogs such as international maritime law codified in the United Nations Convention on the Law of the Sea (1994) and the political status of Antarctica set forth in the Antarctic Treaty (1961). Both are rooted in the notion of the common heritage of humanity, which holds that certain spaces and resources should be held in trust for the benefit of future generations and protected from unilateral exploitation. They also push back against the common law view of property and its use as ``first in time, first in right'' (Buxton, 2004).

In turn, these resource domains are often referred to as ``global commons'' belonging to no individual, corporation or nation. The preamble of the OST recognizes the ``common interest of all mankind in the progress of the exploration and use of outer space for peaceful purposes,'' and that ``the exploration and use of outer space should be carried on for the benefit of all peoples.'' Consistent with the view of space as a commons, the OST’s Article II forbids territorial claims in outer space, and Article IX demands ``appropriate international consultations'' before undertaking activities in space that ``would cause potentially harmful interference with activities of other States Parties in the peaceful exploration and use of outer space''. 

Although the U.S. is a signatory to the OST, current federal policy is "the US. does not view space as a global commons," and it asserts that "Americans should have the right to engage in commercial exploration, recovery, and use of resources in outer space".~\footnote{\href{https://www.federalregister.gov/documents/2020/04/10/2020-07800/encouraging-international-support-for-the-recovery-and-use-of-space-resources}{“Encouraging International Support for the Recovery and Use of Space Resources”} (E.O. 13914 of April 6, 2020).}  But other branches of the federal government, including the military, have considered space to be a commons. The Joint Chiefs of Staff in 2016 issued a statement that the  "[p]rosperity of the United States depends upon its largely uncontested ability to access and use the global commons."~\footnote{\href{https://www.jcs.mil/Portals/36/Documents/Doctrine/concepts/joe_2035_july16.pdf}{“The Joint Force in a Contested and Disordered World”} (Joint Operating Environment 2035, July 14, 2016).} Additionally, President Obama's Secretary of Defense, Leon Panetta, stressed the importance of protecting freedom of access to the global commons, including outer space.~\footnote{\href{https://permanent.fdlp.gov/gpo18079/DefenseStrategicGuidance.pdf}{“Sustaining U.S. global leadership: priorities for 21st century defense”} (2012).} 

As the most densely occupied realm of outer space and the only realm in which a continuous human presence currently exists, it is arguable that LEO space in particular is a de facto global commons. As such, it risks suffering a ``tragedy of the commons,'' (Hardin, 1968) in which resource depletion results from individual users acting independently according to their own self-interest when their activities are not restricted or regulated. 

Regarding the environmental aspects of space, NEPA broadly refers to the ``human environment,'' which includes ``the natural and physical environment and the relationship of people with that environment.''~\footnote{40 CFR §1508.14.} Notably, the framers of NEPA placed no strictures on the reach of the human environment, and in fact note that the term is to be ``interpreted comprehensively.'' Some have argued that the human environment extends to the region of space that humans are capable of accessing physically (Venkatesan et al., 2020). If this is so, then the U.S. Environmental Protection Agency (EPA) is already authorized by Congress to regulate LEO space under existing statute despite the FCC categorical exemption (IDA/Mudd Law, 2020). 


\section{Policy argument and recommendations}

LEO is arguably a global commons, and it will be subject to irreversible and multi-faceted harm though the crowding of hundreds of thousands of new objects being launched into this space. We advocate for new policies and a reimagined regulatory framework that may help prevent a rapidly developing tragedy. Only a radical rethinking of space policy might successfully manage an equally radical and ongoing remaking of space and the night sky, a shared commons rooted in conquest and claim rather than communities.

For policy ideas, we look to previous efforts that successfully headed off the most dire consequences of rapid technology development, such as the Partial Test Ban Treaty (1963)~\footnote{Nuclear Test Ban Treaty, July 26, 1963; Treaties and Other International Agreements Series \#5433; General Records of the U.S. Government; Record Group 11; National Archives.} and Montreal Protocol (1987).~\footnote{Charter of the United Nations and statute of the International Court of Justice, Chapter XXVII, Sec. 2a. (Montreal Protocol on Substances that Deplete the Ozone Layer, Montreal, 16 September 1987).
} We draw further inspiration from the conclusions of the Dark and Quiet Skies for Science and Society conference (Walker et al., 2020b) and the final reports from the SATCON2 Community Engagement Working Group (Venkatesan et al., 2021) and Policy Working Group (Green et al., 2021).

Our policy recommendations are intended to further the cause of better stewardship of the LEO environment. Such changes are essential for truly inclusive policies for a diverse range of space actors. These recommendations are also beneficial for industry, business, and commercial actors, because commercial space operators seek regulatory certainty and a LEO environment free from the undue risk of collisions. Individual groups of stakeholders may have differing and even orthogonal ambitions and practices in near-Earth space, but they share the long-term goals of the peaceful, sustainable use of the skies and space for scientific, economic, commercial, and cultural purposes.

The National Space Council recently drew attention to the need for changes to the status quo in order to achieve a sustainable future in space (Sheetz, 2021). In December 2021, the Council released its "United States Space Priorities Framework."~\footnote{\href{https://www.whitehouse.gov/wp-content/uploads/2021/12/United-States-Space-Priorities-Framework-_-December-1-2021.pdf}{https://www.whitehouse.gov/wp-content/uploads/2021/12/United-States-Space-Priorities-Framework-\_-December-1-2021.pdf}.} This framework emphasizes the importance of the U.S. taking a leadership role to better address important issues in the burgeoning space sector and puts forth the vision that the country will "lead the international community in preserving the benefits of space for future generations." Space situational awareness, space traffic coordination, and space debris are mentioned as areas of priority. To accomplish these and other goals, this framework asserts that "the United States will lead in strengthening global governance of space activities'' and "will engage the international community to uphold and strengthen a rules-based international order for space." The recommendations we outline here offer an effective way forward to achieve these goals.

\subsection{Pause new U.S. megaconstellation operations clearances pending results of a comprehensive review of all aspects of impacts on LEO space}
For the reasons outlined in Section 1.3, it is clear that there are many unresolved concerns relating to deploying satellite mega-constellations. A transition from the hypothetical to the concrete is ongoing as we gather more evidence showing the reality of associated harms. There remain serious and potentially irreversible consequences due to the crowding of LEO space that call for an approach supported by strong precaution. Stakeholder consultations to date have been inadequate, politically powerless groups have been left out of the decision-making process, and the history of technology is replete with examples of hazards associated with rapid adoption of poorly understood innovations. To find fair, durable, and long-term solutions to these issues, it is imperative to halt the practice of granting new FCC operations licenses until the problem is properly studied and regulatory certainty is obtained. 

\subsection{Subject LEO launches and operations clearances to NEPA environmental impact assessments}
It has been argued that the FCC ``categorical exemption'' to environmental impact review of space operations runs counter to the implied will of Congress expressed in NEPA. Federal courts have yet to resolve this tension. While Congress could clarify the intent of NEPA with respect to its applicability in space, changing the existing FCC policy does not require an act of Congress. As an executive branch agency, the categorical exemption could be repealed by executive order, provided that the order is consistent with the FCC’s enabling legislation.~\footnote{Communications Act of 1934, 47 U.S.C. 151 \emph{et seq}.} As such, we recommend that space be considered a human environment, and thus subject to environmental impact assessments.

\subsection{Set legal thresholds for interference to optical, infrared, and radio astronomy, and fine operators for failing to meet them}
This recommendation takes its cue from models such as the Clean Air Act of 1963~\footnote{42 U.S.C. §7401 \emph{et seq}.} and Clean Water Act of 1972,~\footnote{33 U.S.C. §§ 1251–1387.} and their amendments, which tasked the EPA with establishing pollution standards that require reductions in emissions of hazardous air and water pollutants. Congress should conclude that the value of public investment in the American astronomy enterprise, as well as the continued accessibility of LEO space makes both worthy of protection against the effects of large numbers of satellites and take appropriate action as required to ensure that protection. Such an approach is a compromise because it would allow for reasonable utilization of LEO space, much as the Clean Air and Clean Water Acts allow uses that yield some air and water pollution. Activities that result in pollution exceeding legal limits would be deemed unlawful. Financial consequences for failing to comply may nudge satellite operators in the direction of mitigation, whether that is launching fewer objects, launching into lower orbits, darkening objects, or other methods.

\subsection{Require that satellite operators proactively justify the number of objects in their constellations}
The available evidence clearly identifies the number of objects in LEO as a hazard for ground- and space-based OIR astronomy, RFI, and debris generation. Commercial satellite operators could achieve similar business goals with fewer objects. This should be incentivized in the federal review process with preference given to operations clearance requests that entail the fewest number of objects to be launched. Alternately, insofar as most commercial launches take place from U.S. territory, Congress could take the more aggressive step of simply limiting the total number of objects it will permit to be launched. 

\subsection{Create a new oversight agency for LEO space activities}
Existing regulatory bodies involved in satellite issues were not created specifically to address space activities and have many other concerns. Given the increasing pace of launches, as well as the potential of such activity to impact the environment and impair long-term sustainable use of LEO, Congress should mandate the establishment of a new regulatory entity dedicated to  civilian space. Such an agency must be freestanding outside of the FAA and FCC, while working in cooperation with those agencies. We recommend comprehensive legislation consolidating oversight of the country’s civilian space activities into a single agency tasked with implementing U.S. space policy and obtaining compliance both with NEPA and U.S. obligations under the OST.~\footnote{Some authors argue that the United States recognizing ownership of space resources under, e.g., the Commercial Space Launch Competitiveness Act of 2015, is an implicit act of sovereignty that violates U.S. obligations pursuant to the Outer Space Treaty. See, e.g., Su (2017) and Koch (2018).}

Our model for the creation of this agency is the 1970 reorganization of government approved by Congress that established the EPA in the absence of an enabling act.~\footnote{Reorganization Plan No. 3 of 1970 (35 FR 15623, 84 Stat. 2086).} In effecting a similar reorganization, Congress could assign a new agency the launch regulation powers currently delegated to the FAA and the operations authority that the FCC claims under the Communications Act of 1934.~\footnote{The authority for this is Pub. L. 89–554, 80 Stat. 378 (1966), which created 5 U.S.C.} Just as the creation of the EPA recognized a need for effective, centralized oversight of a federal policy realm whose component pieces were scattered across government, so too would the creation of a new agency better serve the interests of stakeholders.

\subsection{Enable space exploration rooted in cultural competency}
The acknowledgment of space as a global commons must include the appropriate entities to ensure that cultural uses of the night sky are consistently protected. As such, we recommend that any institution working on matters related to space science and exploration must establish a cultural ethics and protocol office, or at the very least, identify one or more cultural ethics liaisons. Such offices would review actions taken regarding space and the night sky and ensure that such actions are undertaken in a manner that is respectful to the night sky and its related, diverse cultural practices and traditions. In the United States, where most branches of the federal government have an office for tribal liaisons, we recommend a long-overdue Office of Indigenous Affairs or Office of Tribal Relations at NASA.

Additionally, nations and institutions will benefit from a cultural ethics office or liaison(s) by devising methods to include historically excluded communities in the institution’s work in a respectful manner in fields where these communities are otherwise poorly represented. This will help  ensure that the voices of these historically excluded communities are present in decisions regarding space science, exploration, and commerce.

\subsection{Press for a kinetic ASAT test ban treaty}
Lastly, we argue that the potential for both runaway space debris generation and geopolitical instability resulting from tests of military weapons in space calls for immediate action. The United States should press for a Kinetic Anti-satellite (ASAT) Test Ban Treaty. These tests in space are the above-ground nuclear tests of our time, in that they result in the generation of hazardous debris that crosses all jurisdictional boundaries and threatens the ongoing peaceful use of outer space guaranteed to all OST parties. Nations have come together before (e.g., the Partial Test Ban Treaty of 1963) in agreement that such activities threaten the well-being of all and can be effectively reduced and monitored only through international cooperation.


\section{Conclusions and future directions}

Considering both the current and anticipated  hazards associated with the ongoing crowding of LEO space by satellite mega-constellations, it is essential to proceed cautiously and intentionally in order to ensure the long-term sustainability of this shared resource. While it is possible that significant new regulation of space in one country will simply prompt commercial space operators to relocate their activities to more industry-friendly venues with fewer regulatory hurdles, the U.S. remains a lucrative market for satellite communications, and access to that market is essential to operators’ business plans. Regulation, therefore, remains a powerful tool that may influence future conditions in LEO space. We advocate for a radical revision of the current regulatory patchwork of siloed concerns, narrow legal scope, and more gaps than protections toward a more integrative policy-regulatory framework addressing the many concerns we have raised here. 

We have also drawn attention throughout this article to a number of issues in the current landscape of space law and policy that urgently need clarification from experts and governing bodies. There are numerous areas in the legal and policy arenas that need work, particularly in connection to Indigenous nations. These include:

\begin{itemize}
\item Whether predictions about the contribution of satellites to night sky brightness are correct;
\item How night sky impacts vary according to the number of satellites, their orbital heights and distributions;
\item Whether any particular ``carrying capacity'' of satellites in LEO exists;
\item Which satellite designs are effective at reducing or eliminating their impacts on the visibility of the night sky;
\item Legal obligations for state and private actors in space, given existing treaties between sovereign Indigenous nations and individual states or the Crown;
\item How such treaties can be a part of developing inclusive ethical models of space exploration that honor Indigenous rights and perspectives (Neilson and Ćirković, 2021);
\item The interplay in space law and space policy between individual treaties with Indigenous nations and space treaties such as the OST;
\item The question of whether declaring space as a global commons conflicts with sovereign Indigenous nations’ legal interests and claims as agreed upon in their treaties; 
\item The need for written agreements between industry, spacefaring countries and Indigenous nations that respect these treaties and these communities' sovereignty; 
\item The need to formalize in legal and regulatory structures the protection of the night sky for astronomical research and the progress of science; and
\item Broad national and international discussion and agreement about the consideration of LEO space as part of the environment that should therefore be subject to existing environmental regulatory frameworks.
\end{itemize}

We believe that the night sky belongs to all people, and that space is an ancestral global commons that contains the heritage and future of humanity’s scientific and cultural practices including the sky traditions of Indigenous communities and historically marginalized groups worldwide. As we chart a course for our species in the rest of this century and beyond, the choices we make today for LEO and near-Earth space will set precedents for how they will be used for decades, and perhaps even centuries, to come. It is not just the planet and our species that face an existential crossroads, but our collective future in space. Now is the time to make sensible policy changes as we reimagine and co-create ethical, inclusive, and sustainable approaches for our shared journey in space.


\section{References}
\small

\noindent Arms Control Association (ACA). (2021). \href{https://www.armscontrol.org/act/2021-12/news/russian-asat-test-creates-massive-debris}{\emph{Russian ASAT Test Creates Massive Debris}}.\\

\noindent Balmforth, T. (2021, November 17). \href{https://www.reuters.com/business/aerospacedefense/russia-dismisses-us-criticism-anti-satellite-weapons-test-2021-11-16/}{\emph{'Razor-sharp precision': Russia hails anti-satellite weapons test.}} Reuters.\\

\noindent Bassa, C. G., Hainaut, O. R., \& Galadí-Enríquez, D. (2022). Analytical simulations of the effect of satellite constellations on optical and near-infrared observations. \emph{Astronomy \& Astrophysics}, 657, 1-19. \href{https://doi.org/10.1051/00046361/202142101}{https://doi.org/10.1051/00046361/202142101}\\

\noindent Boley, A. C., \& Byers, M. (2021). Satellite mega-constellations create risks in Low Earth Orbit, the atmosphere and on Earth. \emph{Scientific Reports}, 11(1), 1-8. \href{https://doi.org/10.1038/s41598-021-89909-7}{https://doi.org/10.1038/s41598-021-89909-7}\\ 

\noindent Brown, S., \& Senn, G. (1960). Project SCORE. \emph{Proceedings of the IRE}, 48(4), 624--630. \href{https://doi.org/10.1109/jrproc.1960.287438}{https://doi.org/10.1109/jrproc.1960.287438}\\

\noindent Brown, T. M. (2016). Using light to tell the time of day: sensory coding in the mammalian circadian visual network. \emph{Journal of Experimental Biology}, 219(12), 1779--1792. \href{https://doi.org/10.1242/jeb.132167}{https://doi.org/10.1242/jeb.132167}\\ 

\noindent Bugos, S. (2021, December 1). \href{https://www.armscontrol.org/act/2021-12/news/russian-asat-testcreates-massive-debris}{\emph{Russian ASAT test creates massive debris.}} Arms Control Today.\\

\noindent Buxton, C. R. (2004). Property in outer space: The common heritage of mankind principle vs. the first in time, first in right, rule of property. \emph{Journal of Air Law and Commerce}, 69(4), 689-707. No doi. \href{https://scholar.smu.edu/jalc/vol69/iss4/3}{https://scholar.smu.edu/jalc/vol69/iss4/3}\\

\noindent Clean Air Act, 42 U.S.C. §7401 \emph{et seq}. (1963).\\

\noindent Cole, R. E. (2020). Measurement of the brightness of the Starlink spacecraft named “DARKSAT.” \emph{Research Notes of the American Astronomical Society}, 4(3) 42. \href{https://doi.org/10.3847/2515-5172/ab8234}{https://doi.org/10.3847/2515-5172/ab8234}\\

\noindent Combrinck, W. L., West, M. E., \& Gaylard, M. J. (1994). Coexisting with GLONASS: Observing the 1612 MHz hydroxyl line. \emph{Publications of the Astronomical Society of the Pacific}, 106,807-812. \href{https://doi.org/10.1086/133444}{https://doi.org/10.1086/133444}\\

\noindent Dacke, M., Baird, E., el Jundi, B., Warrant, E. J., \& Byrne, M. (2021). How dung beetles steer straight. \emph{Annual Review of Entomology}, 66(1), 243--256. \href{https://doi.org/10.1146/annurev-ento-042020-102149}{https://doi.org/10.1146/annurev-ento-042020-102149}\\

\noindent Dallas, J. A., Raval, S., Alvarez Gaitan, J. P., Saydam, S., \& Dempster, A. G. (2020). The environmental impact of emissions from space launches: A comprehensive review. \emph{Journal of Cleaner Production}, 255, 120209. \href{https://doi.org/10.1016/j.jclepro.2020.120209}{https://doi.org/10.1016/j.jclepro.2020.120209}\\

\noindent Dark and Quiet Skies Working Groups. (2022). Dark and Quiet Skies for Science and Society: Report and recommendations. United Nations Office for Outer Space Affairs. \href{https://www.iau.org/static/publications/dqskies-book-29-12-20.pdf}{https://www.iau.org/static/publications/dqskies-book-29-12-20.pdf}\\

\noindent De Lucia, V., \& Iavicoli, V. (2019). From Outer Space to Ocean Depths: The Spacecraft Cemetery and the Protection of the Marine Environment in Areas beyond National Jurisdiction. \emph{California Western International Law Journal}, 49(2), 345-389. No doi. \href{https://scholarlycommons.law.cwsl.edu/cwilj/vol49/iss2/4}{https://scholarlycommons.law.cwsl.edu/cwilj/vol49/iss2/4}\\

\noindent Department of Defense. (2012, January). \href{https://permanent.fdlp.gov/gpo18079/DefenseStrategicGuidance.pdf }{Sustaining U.S. Global Leadership: Priorities for 21st Century Defense}.\\

\noindent Dickens, P., \& Ormrod, J. S. (2007). Outer Space and Internal Nature: Towards a Sociology of the Universe. \emph{Sociology}, 41(4), 609--626. \href{https://doi.org/10.1177/0038038507078915}{https://doi.org/10.1177/0038038507078915}\\

\noindent Emlen, S. T. (1970). Celestial Rotation: Its Importance in the Development of Migratory Orientation. \emph{Science}, 170(3963), 1198--1201). \href{https://doi.org/10.1126/science.170.3963.1198}{https://doi.org/10.1126/science.170.3963.1198}\\

\noindent European Space Agency (ESA). (2021). \href{https://www.esa.int/Safety_Security/Clean_Space/How_many_space_debris_ objects_are_currently_in_orbit.}{\emph{How many space debris objects are currently in orbit?}}\\

\noindent \href{https://www.federalregister.gov/documents/2020/04/10/202007800/encouraging-international-support-for-the-recovery-and-use-of-spaceresources}{Executive Order No. 13914}, 85 F.R. 20381 (2020). \\

\noindent Federal Communications Commission (FCC). (n.d.) \href{https://www.fcc.gov/general/satellite}{https://www.fcc.gov/general/satellite}\\

\noindent Federal Water Pollution Control Act, 33 U.S.C. §§ 1251--1387 (1972)\\

\noindent Elizabeth Berlin, Exec.~Deputy Comm., New York State Office of Temporary and Disability Assistance (Aug. 18, 2011). \href{https://www.foodpolitics.com/wp-content/uploads/SNAP-Waiver-Request-Decision.pdf}{https://www.foodpolitics.com/wp-content/uploads/SNAP-Waiver-Request-Decision.pdf}\\

\noindent Foster, R. (2021). Fundamentals of circadian entrainment by light. \emph{Lighting Research Technology}, 53(5), 377--393. \href{https://doi.org/10.1177/14771535211014792}{https://doi.org/10.1177/14771535211014792}\\

\noindent Foust, J. (2022, February 18). \href{https://spacenews.com/russian-asat-debris-creatingsqualls-of-close-approaches-with-satellites/}{\emph{Russian ASAT debris creating “squalls” of close approaches with satellites}.} SpaceNews.\\

\noindent Freeman, R. H. (2020). \href{https://opsjournal.org/DocumentLibrary/Uploads/JSOC_Q@_Draft2_JK_final.pdf}{Overview: Satellite Constellations}. \emph{Journal of Space Operations \& Communicator}, 17(2). No doi. \\

\noindent Gilbert, A., \& Vidaurri, M. (2021). Major Federal Actions Significantly Affecting the Quality of the Space Environment: Applying NEPA to Federal and Federally Authorized Outer Space Activities. \emph{Environs: Environmental Law and Policy Journal}, 44(2), 233-271. No doi. \href{https://environs.law.ucdavis.edu/volumes/44/2/Gilbert.pdf}{https://environs.law.ucdavis.edu/volumes/44/2/Gilbert.pdf}\\

\noindent Govindasamy, B., \& Caldeira, K. (2000). Geoengineering Earth’s radiation balance to mitigate CO2-induced climate change. \emph{Geophysical Research Letters}, 27(14), 2141--2144. \href{https://doi.org/10.1029/1999gl006086}{https://doi.org/10.1029/1999gl006086}\\

\noindent Green, R., Allen, L., Andrade, L., Bohnlein, J., Boley, A., Cooper, P., Dunn, M., Falle, A., Grunsfeld, J., Hanlon, M., Hartley, R. Hofer, C., Jakhu, R., Jansson, G., Jones, T., Knox, D., Krafton, K., Liszt, H., Mishra, N., Mudd, C., Parriott, J., Rosenberg, E., Puxley, P., Raval, V., Rotola, G., Sedwick, R., Simon-Butler, A., Smith, J., Vanotti, M., Walker, C., Williams, A. (2021). SATCON2: Policy Working Group Report. In \href{https://baas.aas.org/pub/q099he5g}{Report of the SATCON2 Workshop}, 12--16 July 2021. \\

\noindent de Gouyon Matignon, L. (2019, November 24). \href{https://www.spacelegalissues.com/whydoes-the-faa-uses-50-miles-for-defining-outer-space/}{\emph{Why does the FAA use 50 miles for defining outer space?}} Space Legal Issues.\\

\noindent Hainaut, O. R., \& Williams, A. P. (2020). Impact of satellite constellations on astronomical observations with ESO telescopes in the visible and infrared domains. \emph{Astronomy \& Astrophysics}, 636, A121. \href{https://doi.org/10.1051/00046361/20203750}{https://doi.org/10.1051/00046361/20203750}\\

\noindent Hall, J., Walker, C., Rawls, M., McDowell, J., Seaman, R., Venkatesan, A., Lowenthal, J., Green, R., Krafton, K., Parriott, J. \href{https://noirlab.edu/public/media/archives/techdocs/pdf/techdoc031.pdf}{Report of the SATCON2 Workshop 1216 July 2021, Executive Summary}. NSF’s NOIRLab. \\

\noindent Hall, S. (2019, June 1). \href{https://www.nytimes.com/2019/06/01/science/starlink-spacexastronomers.html}{After SpaceX Starlink Launch, a Fear of Satellites That Outnumber
All Visible Stars.} New York Times.\\

\noindent Hamacher, D. W., de Napoli, K., \& Mott, B. (2020). Whitening the Sky: light pollution as a form of cultural genocide. arXiv:2001.11527 [physics.pop-ph]. \href{https://arxiv.org/abs/2001.11527}{https://arxiv.org/abs/2001.11527}\\

\noindent Hansen, J. (1994). The big balloon. \emph{Air and Space}, 9(1), 70--77. No doi.\\

\noindent Hardin, G. (1968). The Tragedy of the Commons. \emph{Science}, 162(3859), 1243--1248. \href{https://doi.org/10.1126/science.162.3859.1243}{https://doi.org/10.1126/science.162.3859.1243}\\

\noindent Human environment, 40 C.F.R. §1508.14 (1970).\\

\noindent Impey, C. (2021, Dec 29). \href{https://thehill.com/opinion/technology/587630-2021-more-spacelaunches-than-any-year-in-history-since-sputnik}{\emph{2021: More space launches than any year in history since Sputnik.}} The Hill.\\

\noindent Ingraham, C. (2021, April 27). \href{https://www.washingtonpost.com/business/2021/04/27/starlink-lightpollution/}{\emph{A proliferation of space junk is blocking our view of the cosmos, research shows.}} Washington Post.\\

\noindent International Collaboration and Competition in Space: Oversight of NASA’s Role and Programs: Hearings before the Senate Space and Science Subcommittee, 117th Cong. 1 (2021) (Testimony of James Bridenstine) \\

\noindent International Dark-Sky Association (IDA) and Mudd Law. (2020, March 9). Comment in Response to PRM: Update to the Regulations Implementing the Procedural Provisions of the National Environmental Policy Act, Docket No. CEQ-2019-0003. \\

\noindent International Telecommunication Union Radiocommunication Sector (ITU-R). (2010). \href{https://www.itu.int/pub/R-REP-RA.21882010}{Power flux-density and e.i.r.p. levels potentially damaging to radio astronomy receivers} (RA.2188). \\

\noindent International Telecommunication Union Radiocommunication Sector (ITU-R). (2022). \href{https://www.itu.int/en/ITUR/information/Pages/default.aspx}{Welcome to ITU-R}. \\

\noindent International Union for the Conservation of Nature (2021, September 7). \href{https://www.iucncongress2020.org/motion/084}{Resolution 084 - Taking action to reduce light pollution}. \\

\noindent Jah, M. (2020). Space Object Behavior Quantification and Assessment for Space Security. In K.-U. Schrogl (Ed.). Handbook of Space Security, pp. 961--984. \href{https://doi.org/10.1007/978-3-030-232108_103}{https://doi.org/10.1007/978-3-030-232108\_103}\\

\noindent Jindal, P., Bharti, M. K., \& Chalia, S. (2018). Environmental Implications of Various Divisions of Rocket Launch Operations-A Review. \emph{International Journal of Research}, 5(15), 496-502. No doi. \href{https://journals.pen2print.org/index.php/ijr/article/view/15073}{https://journals.pen2print.org/index.php/ijr/article/view/15073}\\

\noindent Johnson-Freese, J. (2017). Build on the outer space treaty. \emph{Nature}, 650(7675), 182--184. \href{https://doi.org/10.1038/550182a}{https://doi.org/10.1038/550182a}\\

\noindent Joint Chiefs of Staff. (2016, July 14). \href{https://www.jcs.mil/Portals/36/Documents/Doctrine/concepts/joe_2035_jul y16.pdf}{The Joint Force in a Contested and Disordered World}. \\

\noindent Keith, D. W. (2000). Geoengineering the Climate: History and Perspectives. \emph{Annual Review of Energy and the Environment}, 25(1), 245--284. \href{https://doi.org/10.1146/annurev.energy.25.1.245}{https://doi.org/10.1146/annurev.energy.25.1.245}\\

\noindent Kessler, D. J., \& Cour-Palais, B. G. (1978). Collision frequency of artificial satellites: The creation of a debris belt. \emph{Journal of Geophysical Research}, 83(A6), 2637. \href{https://doi.org/10.1029/ja083ia06p02637}{https://doi.org/10.1029/ja083ia06p02637}\\

\noindent Koch, J. S. (2018). Institutional Framework for the Province of all Mankind: Lessons from the International Seabed Authority for the Governance of Commercial Space Mining. \emph{Astropolitics}, 16(1), 1--27. \href{https://doi.org/10.1080/14777622.2017.1381824}{https://doi.org/10.1080/14777622.2017.1381824}\\

\noindent Kocifaj, M., Kundracik, F., Barentine, J. C., \& Bará, S. (2021). The proliferation of space objects is a rapidly increasing source of artificial night sky brightness. \emph{Monthly Notices of the Royal Astronomical Society: Letters}, 504(1), L40--L44. \href{https://doi.org/10.1093/mnrasl/slab030}{https://doi.org/10.1093/mnrasl/slab030}\\

\noindent Kwan, R. (2021, December 27). \href{https://www.theguardian.com/science/2021/dec/28/china-complains-to-unafter-space-station-is-forced-to-move-to-avoid-starlink-satellites}{\emph{China anger after space station forced to move to avoid Elon Musk Starlink satellites}}. Guardian.\\

\noindent Lawler, S. M., Boley, A., \& Rein, H. (2021). Visibility Predictions for Near-Future Satellite Megaconstellations: Latitudes near 50 Degrees will Experience the Worst Light Pollution. \emph{Astronomical Journal}, 163, 21. \href{https://doi.org/10.3847/1538-3881/ac341b}{https://doi.org/10.3847/1538-3881/ac341b}\\

\noindent Lee, A. S., Maryboy, N., Begay, D., Buck, W., Catricheo, Y., Hamacher, D., Holbrook, J., Kimura, K., Knockwood, C., Painting, T. K., \& Varguez, M. (2020). Indigenous Astronomy: Best Practices and Protocols for Including Indigenous Astronomy in the Planetarium Setting. arXiv:2008.05266 [physics.hist-ph]. \href{https://arxiv.org/pdf/2008.05266.pdf}{https://arxiv.org/pdf/2008.05266.pdf} \\

\noindent LeGates, T. A., Fernandez, D. C., \& Hattar, S. (2014). Light as a central modulator of circadian rhythms, sleep and affect. \emph{Nature Reviews Neuroscience}, 15,7, 443--454. \href{https://doi.org/10.1038/nrn3743}{https://doi.org/10.1038/nrn3743}\\

\noindent Le May, S., Gehly, S., Carter, B. A., \& Flegel, S. (2018). Space debris collision probability analysis for proposed global broadband constellations. \emph{Acta Astronautica}, 151, 445--455. \href{https://doi.org/10.1016/j.actaastro.2018.06.036}{https://doi.org/10.1016/j.actaastro.2018.06.036}\\

\noindent Letter from Samatha Fonder, NASA Representative to the Commercial Space Transportation Interagency Group Space Operations Mission Directorate, Launch Services Office, to Marlene Dortch, Secretary, Federal Communications Commission (February 14, 2022). \href{https://www.scribd.com/document/557924666/NTIA-NASA-NSF-letter-toFCC-regarding-Starlink-Gen-2}{https://www.scribd.com/document/557924666/NTIA-NASA-NSF-letter-toFCC-regarding-Starlink-Gen-2}\\

\noindent Lintott, C., \& Lintott, P. (2020). Satellite megaclusters could fox night-time migrations. \emph{Nature}, 586(7831), 674. \href{https://doi.org/10.1038/d41586-02003007-8}{https://doi.org/10.1038/d41586-02003007-8}\\

\noindent Martin, P. G., Kruk, S., Popescu, M., Merin, B., Mahlke, M., Carry, B., Thomson, J., Karadag, S., Racero, E., Giordano, F., Baines, D., \& De Marchi, G. (2021). Hubble Asteroid Hunter: Identifying Asteroid Trails in Hubble Space Telescope Images. \emph{Bulletin of the American Astronomical Society}, 53(7). \href{https://baas.aas.org/pub/2021n7i101p08}{https://baas.aas.org/pub/2021n7i101p08}\\

\noindent Massatt, P., \& Zeitzew, M. (1998). \href{https://www.ion.org/publications/abstract.cfm?articleID=584}{The GPS Constellation Design - Current and Projected}. \emph{Proceedings of the 1998 National Technical Meeting of The Institute of Navigation, Long Beach, CA, January 1998} (pp. 435--445).\\

\noindent Mauck, B., Gläser, N., Schlosser, W., \& Dehnhardt, G. (2008). Harbour seals (\emph{Phoca vitulina}) can steer by the stars. \emph{Animal Cognition}, 11(4), 715--718. \href{https://doi.org/10.1007/s10071-008-0156-1}{https://doi.org/10.1007/s10071-008-0156-1} \\
 
\noindent McDowell, J. C. (2020). The Low Earth Orbit Satellite Population and Impacts of the SpaceX Starlink Constellation. \emph{The Astrophysical Journal}, 892(2), L36. \href{https://doi.org/10.3847/2041-8213/ab8016}{https://doi.org/10.3847/2041-8213/ab8016}\\

\noindent Milazzo, M., Richey, C., Piatek, J., Vaughan, A., \& Venkatesan, A. (2021). The Growing Digital Divide and its Negative Impacts on NASA’s Future Workforce. \emph{Bulletin of the American Astronomical Society}, 53(4). \href{https://doi.org/10.3847/25c2cfeb.20f73bfd}{https://doi.org/10.3847/25c2cfeb.20f73bfd}\\

\noindent Mitigation of Orbital Debris in the New Space Age, 85 F.R. 52422 (2020).\\

\noindent Mouritsen, H., \& Larsen, O. N. (2001). Migrating songbirds tested in computer controlled Emlen funnels use stellar cues for a time-independent compass. \emph{Journal of Experimental Biology}, 204(22), 3855--3865. \href{https://doi.org/10.1242/jeb.204.22.3855}{https://doi.org/10.1242/jeb.204.22.3855}\\

\noindent National Environmental Policy Act of 1969, 42 U.S.C. §§4321 \emph{et seq}.\\

\noindent National Space Council (NSC). (2021, December). \href{https://www.whitehouse.gov/wpcontent/uploads/2021/12/United-States-Space-Priorities-Framework-_-December-1-2021.pdf}{United States Space Priorities Framework}.\\

\noindent Neilson, H., \& Ćirković, E. (2021). Indigenous rights, peoples, and space exploration: A response to the Canadian Space Agency (CSA) Consulting Canadians on a framework for future space exploration activities. arXiv:2104.07118 [astro-ph.IM]. \href{https://arxiv.org/abs/2104.07118}{https://arxiv.org/abs/2104.07118}\\

\noindent Nye, D. E. (2006). Technology matters: Questions to live with. MIT Press.\\

\noindent Pardini, C., \& Anselmo, L. (2019). Uncontrolled re-entries of spacecraft and rocket bodies: A statistical overview over the last decade. \emph{Journal of Space Safety Engineering}, 6,(1), 30--47. \href{https://doi.org/10.1016/j.jsse.2019.02.001}{https://doi.org/10.1016/j.jsse.2019.02.001}\\

\noindent Pelton, J., Sgobba, T., \& Trujillo, M. (2020). Space Safety. In K.-U. Schrogl (Ed.). Handbook of Space Security (pp. 265--298). Springer International Publishing. \href{https://doi.org/10.1007/978-3-030-23210-8_50}{https://doi.org/10.1007/978-3-030-23210-8\_50}\\

\noindent Permanent Mission of China to the United Nations. (2021, December 10). Note verbale dated 3 December 2021 from the Permanent Mission of China to the United Nations (Vienna) addressed to the Secretary-General. A/AC.105/1262. \href{https://www.unoosa.org/oosa/en/oosadoc/data/documents/2021/aac.105/aa c.1051262_0.html}{https://www.unoosa.org/oosa/en/oosadoc/data/documents/2021/aac.105/aa c.1051262\_0.html}\\

\noindent Pultarova, T. (2022, February 7). \emph{International Astronomical Union launches new center to fight satellite megaconstellation threat}. Space.com. \href{https://www.space.com/iau-center-protect-astronomy-megaconstellation-threat}{https://www.space.com/iau-center-protect-astronomy-megaconstellation-threat}\\

\noindent Rawls, M. L., Thiemann, H. B., Chemin, V., Walkowicz, L., Peel, M. W., \& Grange, Y. G. (2020). Satellite Constellation Internet Affordability and Need. \emph{Research Notes of the American Astronomical Society}, 4(10), 189. \href{https://doi.org/10.3847/2515-5172/abc48e}{https://doi.org/10.3847/2515-5172/abc48e}\\

\noindent Reorganization Plan No. 3, 35 F.R. 15623 (1970).\\

\noindent Rosenberg, N. W. (1966). Chemical Releases at High Altitudes. \emph{Science}, 152(3725), 1017--1027. No doi. \href{http://www.jstor.org/stable/1718188}{http://www.jstor.org/stable/1718188}\\

\noindent Rossi, A., Cordelli, A., Farinella, P., \& Anselmo, L. (1994). Collisional evolution of the Earth's orbital debris cloud. \emph{Journal of Geophysical Research: Planets}, 99(E11), 23195-23210. \href{https://doi.org/10.1029/94JE02320}{https://doi.org/10.1029/94JE02320}\\

\noindent Rotola, G., \& Williams, A. (2021). Regulatory Context of Conflicting Uses of Outer Space: Astronomy and Satellite Constellations. \emph{Air and Space Law}, 46(4/5), 545--568. No doi. \href{https://kluwerlawonline.com/journalarticle/Air+and+Space+Law/46.4/AILA 2021031}{https://kluwerlawonline.com/journalarticle/Air+and+Space+Law/46.4/AILA 2021031}\\

\noindent Roulette, J. (2021, November 15). \href{https://www.nytimes.com/2021/11/15/science/russia-anti-satellite-missiletest-debris.html}{\emph{Debris From Test of Russian Antisatellite Weapon Forces Astronauts to Shelter}}. New York Times.\\

\noindent Ryan, R. (2020). The Fault in Our Stars: Challenging the FCC’s Treatment of Commercial Satellites as Categorically Excluded from Review under the National Environmental Policy Act. \emph{Vanderbilt Journal of Entertainment and Technology Law}, 22(4), 923. No doi. \href{https://scholarship.law.vanderbilt.edu/jetlaw/vol22/iss4/8/}{https://scholarship.law.vanderbilt.edu/jetlaw/vol22/iss4/8/}\\

\noindent Sanchez, S. (2022, February 15). \href{https://www.borderreport.com/regions/texas/faadelays-release-of-space-x-evaluation-on-holy-grail-of-space-rockets/}{\emph{FAA delays release of SpaceX evaluation on ‘holy grail’ of space rockets}}. Border Report.\\

\noindent Schulz, L., \& Glassmeier, K.-H. (2021). On the anthropogenic and natural injection of matter into Earth’s atmosphere. \emph{Advances in Space Research}, 67(3), 1002--1025. \href{https://doi.org/10.1016/j.asr.2020.10.036}{https://doi.org/10.1016/j.asr.2020.10.036}\\

\noindent Seitzer, P., \& Tyson, J. A. (2021). \href{https://conference.sdo.esoc.esa.int/proceedings/sdc8/paper/112/SDC8paper112.pdf}{Large LEO constellations, astronomy, and space debris mitigation}. \emph{Proceedings of the 8th European Conference on Space Debris}. 8th European Conference on Space Debris, Darmstadt, Germany. No doi. \\

\noindent Sheetz, M. (2021, December 1). \emph{White House unveils ‘Space Priorities Framework’ as VP Kamala Harris leads first space council meeting}. CNBC. \href{https://www.cnbc.com/2021/12/01/white-house-vp-kamala-harris-unveilsspace-priorities-framework.html}{https://www.cnbc.com/2021/12/01/white-house-vp-kamala-harris-unveilsspace-priorities-framework.html}\\

\noindent Siddiqui, F. (2021, November 8). \emph{Tesla’s recent Full Self-Driving update made cars go haywire. It may be the excuse regulators needed}. Washington Post. \href{https://www.washingtonpost.com/technology/2021/11/08/tesla-regulationelon-musk/}{https://www.washingtonpost.com/technology/2021/11/08/tesla-regulationelon-musk/}\\

\noindent Solon, O. (2021, December 8). \href{https://www.cnbc.com/2021/12/08/disgruntled-neighbors-and-dwindlingshorebirds-jeopardize-spacex-expansion.html}{\emph{Disgruntled neighbors and dwindling shorebirds jeopardize SpaceX expansion}}. NBC News. \\

\noindent Soto, M. (1996). \href{https://nsuworks.nova.edu/ilsajournal/vol3/iss1/10}{General Principles of International Environmental Law}. \emph{ILSA Journal of International \& Comparative Law}, 3(1), pp. 194-209. No doi. \\

\noindent Su, J. (2017). Legality of unilateral exploitation of space resources under international law. \emph{International and Comparative Law Quarterly}, 66(4), 991--1008. \href{https://doi.org/10.1017/s0020589317000367}{https://doi.org/10.1017/s0020589317000367}\\

\noindent Thiele, S., \& Boley, A. (2021). Investigating the risks of debris-generating ASAT tests in the presence of megaconstellations. arXiv:2111.12196 [astro-ph.EP]. \href{http://arxiv.org/abs/2111.12196}{http://arxiv.org/abs/2111.12196}\\

\noindent Tran, K., Lim, D., Min, S., \& Mavris, D. N. (2018). Noise and Sonic Boom Analysis from Rocket Launches. 2018 AIAA/CEAS Aeroacoustics Conference. \href{https://doi.org/10.2514/6.2018-2816}{https://doi.org/10.2514/6.2018-2816}\\

\noindent Tregloan-Reed, J., Otarola, A., Ortiz, E., Molina, V., Anais, J., González, R., Colque, J. P., \& Unda-Sanzana, E. (2020). First observations and magnitude measurement of Starlink’s Darksat. \emph{Astronomy \& Astrophysics}, 637, L1. \href{https://doi.org/10.1051/0004-6361/202037958}{https://doi.org/10.1051/0004-6361/202037958}\\

\noindent Tyson, J. A., Ivezić, Ž., Bradshaw, A., Rawls, M. L., Xin, B., Yoachim, P., Parejko, J., Greene, J., Sholl, M., Abbott, T. M. C., \& Polin, D. (2020). Mitigation of LEO Satellite Brightness and Trail Effects on the Rubin Observatory LSST. \emph{Astronomical Journal}, 160(5), 226. \href{https://doi.org/10.3847/1538-3881/abba3e}{https://doi.org/10.3847/1538-3881/abba3e}\\

\noindent Union of Concerned Scientists (UCS). (2021, September 1). \href{https://www.ucsusa.org/resources/satellite-database}{UCS Satellite Database}.\\

\noindent \href{https://law.justia.com/cases/federal/appellate-courts/cadc/181129/18-1129-2019-08-09.html}{\emph{United Keetoowah Band of Cherokee Indians in Oklahoma v. FCC}}, No. 18-1129 (D.C. Cir. 2019). \\

\noindent United Nations (UN). (1959, December 1). The Antarctic Treaty, 12 U.S.T. 794, 402 U.N.T.S. 71, 19 I.L.M 860.\\

\noindent United Nations (UN). (1963, August 5). Treaty Banning Nuclear Weapons in the Atmosphere, in Outer Space and Under Water, 14 U.S.T. 1313, 480 U.N.T.S. 43, 2 I.L.M. 889.\\

\noindent United Nations (UN). (1967, January 27). Treaty on Principles Governing the Activities of States in the Exploration and Use of Outer Space, including the Moon and Other Celestial Bodies, 18 U.S.T. 2410, 610 U.N.T.S. 205, 61 I.L.M. 386.\\

\noindent United Nations (UN). (1972, March 29). Convention on International Liability for Damage Caused by Space Objects, 24 U.S.T. 2389, 861 U.N.T.S. 187, 10 I.L.M. 965.\\

\noindent United Nations (UN). (1982, December 10). Convention on the Law of the Sea, 1833 U.N.T.S. 397, 21 I.L.M. 1261.\\

\noindent United Nations (UN). (1987, September 16). Montreal Protocol on Substances that Deplete the Ozone Layer, 1522 U.N.T.S. 3, 26 I.L.M. 1541.\\

\noindent U.S. Commercial Space Launch Competitiveness Act, P.L. 114-90 (2015)\\

\noindent Venkatesan, A., Begay, D., Burgasser, A. J., Hawkins, I., Kimura, K. I., Maryboy, N., \& Peticolas, L. (2019). Towards inclusive practices with Indigenous knowledge. \emph{Nature Astronomy}, 3(12), 1035-1037. \href{https://www.nature.com/articles/s41550-019-0953-2}{https://www.nature.com/articles/s41550-019-0953-2}\\

\noindent Venkatesan, A., Lowenthal, J., Prem, P., \& Vidaurri, M. (2020). The impact of satellite constellations on space as an ancestral global commons. \emph{Nature Astronomy}, 4(11), 1043--1048. \href{https://doi.org/10.1038/s41550-020-01238-3}{https://doi.org/10.1038/s41550-020-01238-3}\\

\noindent Venkatesan, A., et al.~(2021). Perspectives from Indigenous Communities. \href{https://noirlab.edu/public/media/archives/techdocs/pdf/techdoc033.pdf}{SATCON2 Working Group Reports}, eds. Hall, J. and Walker, C. NOIRLab. \\

\noindent Vera Rubin Observatory (VRO). (2020, May 19). \href{https://www.lsst.org/content/lsst-statement-regarding-increased-deploymentsatellite-constellations}{Impact of Satellite Constellations}\\

\noindent \href{https://dockets.justia.com/docket/circuit-courts/cadc/21-1123}{\emph{Viasat, Inc., v. FCC and Space Exploration Holdings, LLC}}, No. 21-1123 (D.C. Cir. 2021).\\

\noindent Voelsen, D. \& Stiftung Wissenschaft Und Politik. (2021, April). Internet from space. \emph{German Institute for International and Security Affairs}. \href{https://doi.org/10.18449/2021RP03}{https://doi.org/10.18449/2021RP03}\\

\noindent Walker, C., Hall, J., Allen, L., Green, R. , Seitzer, P. , Tyson, A. , Bauer, A. , Krafton, K., Lowenthal, J., Parriott, J. , Puxley, P., Abbott, T., Bakos, G. , Barentine, J., Bassa, C., Blakeslee, J. , Bradshaw, A., Cooke, J., Devost, D., Galadí, D., Haase, F., Hainaut, O., Heathcote, S., Jah, M., Krantz, H., Kucharski, D., McDowell, J., Mróz, P., Otarola, A., Pearce, E., Rawls, M., Saunders, C., Seaman, R., Siminski, J., Snyder, A., Storrie-Lombardi, L., Tregloan-Reed, J., Wainscoat, R., Williams, A. \& Yoachim, P. (2020a). \href{https://aas.org/sites/default/files/2020-08/SATCON1-Report.pdf}{Impact of Satellite Constellations on Optical Astronomy and Recommendations Toward Mitigations}. NSF’s NOIRLab. \\

\noindent Walker, C., Di Pippo, S., Aubé, M., Barentine, J., Benkhaldoun, Z., Benvenuti, P., Bouroussis, C., Green, R., Hearnshaw, J., Liszt, H., Lowenthal, J. D., Muñoz Tuñón, C., Nield, K., Ricard, N., Rodriguez Espinosa, J. M., Sanhueza, P., Varela Pérez, A.M., Williams, A. (2020b). \href{https://www.iau.org/static/publications/dqskiesbook-29-12-20.pdf}{Dark and Quiet Skies for Science and Society}. International Astronomical Union. \\

\noindent Walker, C., \& Benvenuti, P. (Eds.) (2022). Dark and Quiet Skies for Science and Society II Working Group Reports. NSF’s NOIRLab. \href{https://www.iau.org/static/science/scientific_bodies/working_groups/286/da rk-quiet-skies-2-working-groups-reports.pdf}{https://www.iau.org/static/science/scientific\_bodies/working\_groups/286/dark-quiet-skies-2-working-groups-reports.pdf}\\

\noindent Witze, A. (2020, Aug 26). \href{https://doi.org/10.1038/d41586-020-02480-5}{\emph{How satellite ‘megaconstellations’ will photobomb astronomy images}}. Nature.\\

\noindent 

\end{document}